# A Comprehensive Review of Pre-Darcy Flows in Low-Permeability Porous Media

Yuntian Teng[a], Zihao Li[b], Cheng Chen[a*]

[a] Department of Civil, Environmental and Ocean Engineering, Stevens Institute of Technology, Hoboken, NJ 07030

[b] Pacific Northwest National Laboratory, Richland, WA 99354, United States

## Abstract

The widely used Darcy's law specifies a linear relation between the Darcy velocity of a fluid flow and the pressure gradient that drives the flow. However, studies have shown that Darcy velocity can exhibit a nonlinear dependence on the pressure gradient in low-permeability porous media such as clay and shale when the pressure gradient is adequately low. This phenomenon is referred to as non-Darcian flow or pre-Darcy flow. This paper provides a comprehensive review of the theories, experimental data, and modeling methods for pre-Darcy flow in low-permeability porous media. This paper begins by outlining the fundamental mechanisms underlying pre-Darcy flow, which regulate the unique characteristics such as non-linear pressure gradients and the dependence on fluid-rock interactions. The paper then proceeds to present a thorough compilation of experimental investigations performed in various low-permeability geomaterials including tight sandstones, shales, and clays. A detailed discussion of the methodologies employed in these experimental studies is provided, which covers different aspects such as core sample preparation, permeability measurement techniques, and threshold pressure gradient measurements. The results and findings from the experiments are discussed, demonstrating the influence of different factors such as pore geometry, fluid type, and pressure conditions on the onset of pre-Darcy flow. In the next step, the paper reviews empirical and theoretical models and simulation methods that have been developed to fit and interpret experimental data. At the end, the review underscores the challenges encountered in conducting and interpreting pre-Darcy flow experiments and suggests future research directions. By compiling and analyzing previous experimental investigations with respect to pre-Darcy flow, the paper aims to offer a valuable resource for researchers and practitioners who seek to enhance their understanding of fluid dynamics in low-permeability geomaterials. This work will provide insights into the application of pre-Darcy flow in numerous natural and engineered processes, such as shale oil and gas recovery, contaminant transport in low-permeability aquifers, and geological disposal of nuclear waste.

* Corresponding author: Cheng Chen (cchen6@stevens.edu)

## 1. Introduction

Identifying the properties of fluid flows in subsurface porous media is critical to many natural and engineered processes, such as petroleum engineering, geothermal energy recovery, geological $CO_2$ storage, subsurface $H_2$ storage, contamination remediation in aquifers, and geological disposal of high-level nuclear waste (Guo et al., 2022; Li et al., 2022; McClure & Horne, 2014; Nadim et al., 2000; Zhao et al., 2023; Zheng et al., 2020). The widely used Darcy's law describes the linear relationship between the hydraulic pressure gradient and the Darcy flow velocity. However, as many studies pointed out, the relationship between the Darcy flow velocity and hydraulic pressure gradient can be nonlinear in low-permeability porous media when the hydraulic pressure gradient is sufficiently low, which is referred to as non-Darcian or pre-Darcy flow (Hansbo, 2001; Kutílek, 1972; Liu et al., 2012; Miller & Low, 1963; Soni et al., 1978). Different from the non-Darcian flow at high flow velocities occurring in near-wellbore regions or in hydraulic fractures (Balhoff & Wheeler, 2009), the non-Darcian flow at low flow velocities usually disturbs the onset of the fluid flow in low-permeability porous media by lowering or nullifying the flow velocity. To eliminate the confusion with non-Darcian flow at high flow velocities, in the rest of this paper non-Darcian flow at low flow velocities is referred to as pre-Darcy flow. In pre-Darcy flow, the hydraulic pressure gradient needs to exceed a certain threshold to trigger the fluid flow. The complexity of pre-Darcy flow dynamics poses significant challenges in accurately modeling and predicting fluid movement in subsurface environments, especially in low-permeability materials where traditional Darcy flow-based models fail to capture the nuances of flow behaviors. The nonlinear relation between the fluid flow velocity and hydraulic pressure gradient at low flow velocities has profound practical implications. For instance, in enhanced oil recovery and geological $CO_2$ sequestration, underestimating the threshold pressure gradient could lead to inefficient fluid injection strategies or unexpected migration patterns (Longmuir, 2004; Siddiqui et al., 2016). Similarly, in the context of groundwater contamination and subsurface nuclear waste disposal, accurately predicting the onset of flows is crucial for assessing the risk of contaminant transport and ensuring the long-term safety of the storage sites, respectively (Liu et al., 2012).

Investigation of the pre-Darcy flow in low-permeability porous media is experimentally challenging, which typically requires specialized experimental equipment and methods to measure ultra-low threshold pressure gradients or flow rates. The growing body of research on pre-Darcy flow has unveiled several underlying mechanisms that contribute to this phenomenon, including the interactions between fluids and solids, microscale heterogeneity and pore structure complexities, viscoelastic effects, electrochemical effects in certain types of media, and surface roughness and wettability (Altmann et al., 2012; Bourg & Ajo-Franklin, 2017; Fan et al., 2020; Siddiqui et al., 2016; Whittaker et al., 2019; Zhang et al., 2022). Additionally, it was observed that factors such as temperature, fluid properties, and geochemical environments can influence the threshold pressure gradient, thereby further complicating the prediction and management of fluid flow in these contexts. Many theoretical models have been developed to characterize pre-Darcy flow, especially in the perspective of the permeability and threshold pressure gradient (Chen, 2019; Hansbo, 1960; Jiang et al., 2012; Wang & Sheng, 2017; Xiong et al., 2017). Most models describe the pre-Darcy flow well in certain conditions and can be applied to large-scale simulations such as geological storage of high-level nuclear waste. These applications of pre-Darcy flow properties usually involve intricate geomaterials such as bentonite, which can work as the buffer and backfill material for subsurface nuclear waste disposal (Tsang et al., 2012). Bentonite or the mixture of bentonite and sand exhibits ultra-low permeability and can swell when in contact with water, which further reduces the permeability. The complicated fluid-solid interactions between bentonite and water cause the pre-Darcy flow, and the swelling hysteresis of bentonite also further complicates the prediction and management of fluid flow in these materials (Teng et al., 2022).

This review aims to consolidate the current understanding of pre-Darcy flow in low-permeability porous media by examining a variety of experimental investigations. We delve into the methodologies employed to characterize pre-Darcy flow, including core flooding experiments and measurements of threshold pressure gradients. We explore how different parameters affect the onset and behaviors of pre-Darcy flow, thereby drawing insights from recent studies across diverse geological settings. Furthermore, we summarize the existing models describing nonlinear pre-Darcy behaviors and examine the models using experimentally-measured data. By synthesizing these findings, this review not only provides a comprehensive overview of the state-of-the-art in pre-Darcy flow research but also identifies knowledge gaps and potential avenues for future explorations. Such insights are invaluable for improving fluid flow models, thereby enhancing the efficiency of subsurface operations and mitigating environmental risks associated with fluid movements in low-permeability media. In the subsequent sections, we will systematically discuss the experimental methods and findings, the theoretical advancements informed by these experiments, and the practical implications of pre-Darcy flow in various hydrogeological applications. This comprehensive perspective aims to bridge the gap between experimental observations and practical applications, thereby fostering a deeper understanding of fluid dynamics in low-permeability porous media.

## 2. Fundamentals of Pre-Darcy Flow

### 2.1. Darcy's law

Darcy's Law is a fundamental equation that describes the flow of a fluid through a porous medium. Formulated by Henry Darcy in the mid-19th century, it was originally developed to understand the flow of water through sand filters in fountains (Darcy, 1856). Today, Darcy's Law is a cornerstone in the fields of hydrogeology, petroleum engineering, and environmental engineering (Brown, 2002). Darcy's Law states that the flow rate of a fluid through a porous medium is proportional to the gradient of the hydraulic head (i.e., hydraulic pressure difference) and the permeability of the medium:

$$q = -\frac{kA}{\mu}\frac{\Delta P}{L} \qquad (1)$$

where $q$ is the flow rate of the fluid ($m^3/s$), $k$ is the permeability ($m^2$), $A$ is the cross-sectional area of the porous medium ($m^2$), $\Delta P$ is the pressure difference across the flow direction (Pa), $\mu$ is the fluid viscosity (Pa·s), and $L$ is the length of the porous medium in the flow direction (m). The negative sign indicates that the flow of water occurs from higher pressure to lower pressure.

Although the Darcy's law provides a fundamental framework to understand fluid flow in porous media, it is based on assumptions that the incompressible Newtonian fluid follows the laminar flow characteristics. However, in real-world scenarios, deviations from these assumptions necessitate adjustments or the use of more complicated flow models.

### 2.2. Pre-Darcy flow

Pre-Darcy flow refers to the fluid flow regime in porous media that deviates from the linear behavior described by the Darcy's law. It is typically observed at low flow velocities in low-permeability geomaterials such as tight sandstones, shales, and clays. In the pre-Darcy flow regime, the fluid flow needs to be triggered by overcoming a certain hydraulic pressure gradient; therefore, the flow rate does not increase proportionally with the increase of the hydraulic pressure gradient. In general, the threshold pressure gradient is a key parameter to determine the pre-Darcy flow properties (Prada & Civan, 1999). **Figure 1** shows the difference between regular Darcy flow and pre-Darcy flow, as well as the definition of the true and extrapolated threshold pressure gradients. Compared to the Darcy's law, which is a linear relation, fluid flow in the pre-Darcy flow regime is not triggered until the hydraulic pressure gradient exceeds the true threshold pressure gradient, $J_t$. Beyond $J_t$, as the pressure gradient increases, the flow regime gradually transitions from pre-Darcy to Darcy. This transition is not abrupt but rather an intermediate stage where the dependence of the flow rate on the pressure gradient gradually transitions from nonlinear to linear. This behavior is indicative of that the fluid pathways become fully mobilized, thereby allowing for a more linear relationship between the flow rate and pressure gradient. In many situations, the extrapolated threshold pressure gradient, $J_c$, is widely used to simplify the processes of experimental measurements and numerical modeling because direct measurements of $J_t$ in the laboratory is time consuming and subject to various challenges caused by the ultra-low flow rate and pressure gradient. The discrepancy between $J_t$ and $J_c$ necessitates careful consideration in experimental designs and data interpretation, especially when characterizing the flow properties in unconventional reservoirs or assessing the migration of contaminants in low-permeability aquifers.

It is important to note that $J_t$ is not merely a static value but can vary based on factors such as fluid viscosity, temperature, and the specific characteristics of the porous medium, including its surface wettability and chemical composition. Understanding pre-Darcy flow is not only critical to hydrogeological and petroleum applications but also provides insights into other fields such as environmental engineering and soil science. The accurate characterization of pre-Darcy flow impacts the efficiency of resource extraction, the effectiveness of groundwater remediation, and the reliability of geological carbon storage.

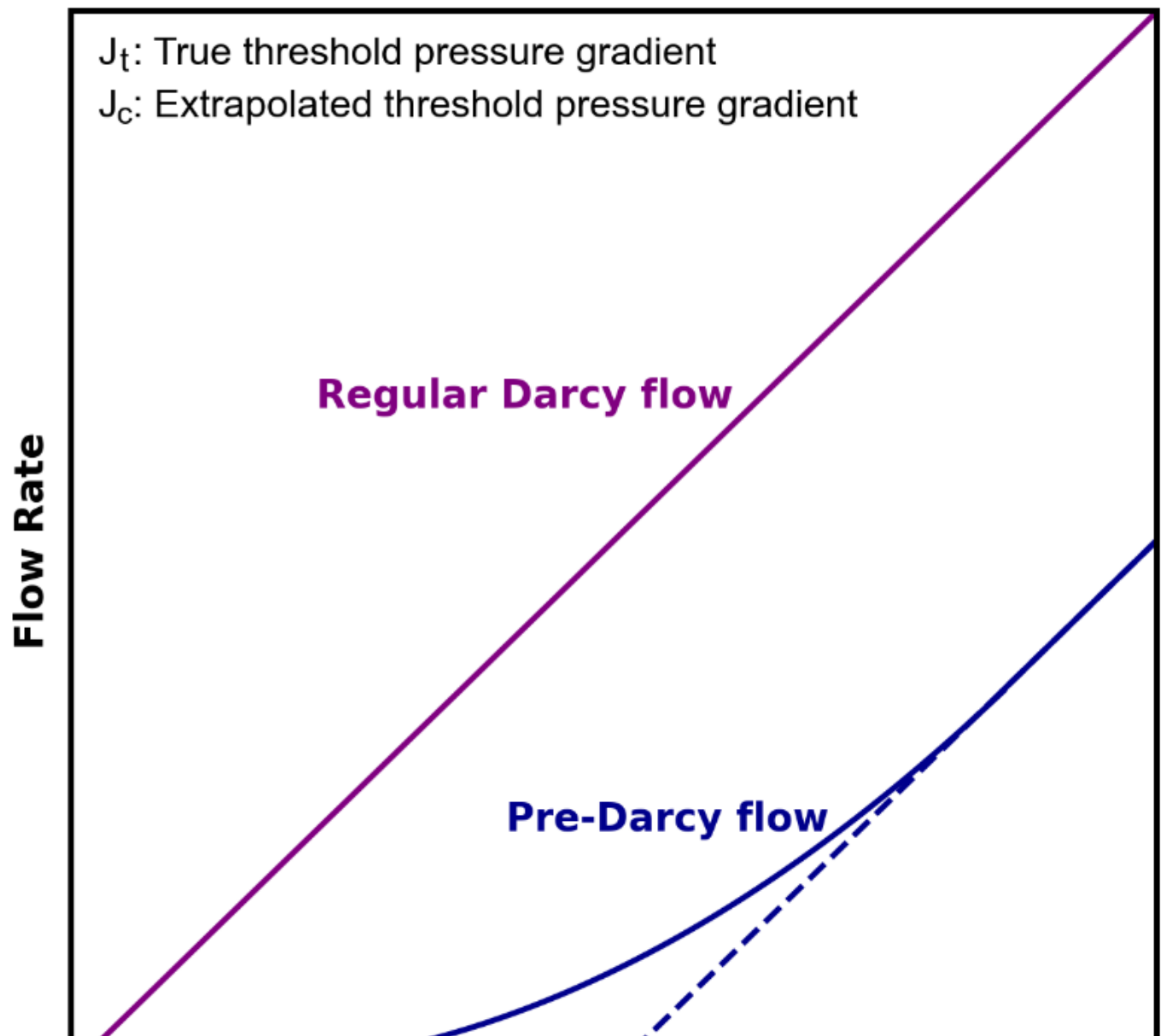

**Figure 1.** Comparison between regular Darcy flow and pre-Darcy flow. The pre-Darcy flow regime starts from the true threshold pressure gradient, $J_t$. An extrapolated threshold pressure gradient, $J_c$, is also widely used in many studies to simplify the experimental measurements.

## 3. Pre-Darcy Mechanisms in Low-Permeability Porous Media

Low-permeability geomaterials such as shales, tight sandstones, and clays, are characterized by their limited pore spaces and low connectivity among these pores. This results in significantly reduced fluid flow rates through these materials compared to more permeable rocks. Fluid flow in such low-permeability porous media usually involves pre-Darcy flow. Shales are fine-grained sedimentary rocks composed of a mix of clay minerals and other minor constituents like quartz and calcite. They typically form from the compaction of silt and clay-sized particles in low-energy environments like deep marine settings. The ultrafine structure of shales significantly restricts fluid flow, and the sedimentary layered structures result in the anisotropy of permeability. Tight sandstones are sandstone reservoirs with very low permeability, which are made primarily of sand grains, often coated with clay or cemented by minerals such as quartz and calcite. The reduced space between grains and pore-throat blockage by cementation are the primary reasons for the low permeability. Clays consist of fine-grained minerals with a sheet-like structure, offering a very high surface area to volume ratio. Clays can absorb water and swell, further reducing the porosity and permeability. Clay minerals are a major component of soils and are important in various geological processes. These materials often act as cap rocks in petroleum geology and carbon dioxide geological storage, trapping hydrocarbons and stored carbon dioxide in more permeable rocks underneath to prevent leakage and transport. Clay-rich shales are sources of unconventional hydrocarbons, such as shale oil and gas, which have become increasingly important as energy resources. Clays and clayey geomaterials also work as the backfill material at geological disposal sites of high-level nuclear waste because of their ultra-low permeability, high swelling capacity, and strong retardation of radionuclide transport. Therefore, understanding the flow properties, especially the pre-Darcy flow properties of these materials, is crucial, particularly in the context of swelling clays.

In general, shale and tight sandstone have more similar properties with respect to the pre-Darcy flow behavior compared to clays. The mechanisms causing the nonlinear flow behavior at low hydraulic pressure gradients and low flow velocities in shales and tight sandstones were summarized in **Table 1**. In clays, some of the mechanisms summarized in Table 1 also affect the nonlinear flow behavior, such as the complex structures, surface effects, viscoelastic flows, and electrokinetic effects. However, more specific mechanisms causing the nonlinear flow behavior in clays are unique and we summarized them in **Table 2**. We included the threshold pressure gradient as one of the mechanisms in clays because the strong capillary forces and electrostatic interactions within the fine pores of the clay matrix are often considered major causes of the immobile water layer, which is important in constructing theoretical models of pre-Darcy flows in clays. The reduction of the immobile water layer thickness is regarded as one dominating parameter in such models.

**Table 1.** Mechanisms causing pre-Darcy flow in shales and tight sandstones.

| | **Shale** | **Tight Sandstone** |
|---|---|---|
| **Knudsen diffusion and slip flow (Klinkenberg effect)** | In extremely fine pores of shales and tight sandstones, especially those with pore diameters less than the mean free path length of the fluid molecules, Knudsen diffusion becomes significant. The movement of gas molecules is more influenced by collisions with pore walls than with other molecules, leading to a flow behavior that deviates from conventional bulk fluid flow. This effect is particularly relevant to gases and a key factor for gas transport in shales and tight sandstones. In the nanopores of shales and in the small pore throats of tight sandstones, especially under low-pressure conditions, gas molecules may slip along the pore walls instead of adhering to them. This slippage increases the effective permeability for gas flow, thereby deviating from the Darcy's law (Cui, 2019; Li et al., 2020; Moghadam & Chalaturnyk, | |
| | 2017; Swami et al., 2012; W. Zhang et al., 2020). | |
| **Complex microstructures** | Shales possess a complex microstructure with a wide range of pore sizes, including nanopores. The distribution and connectivity of these pores significantly impact fluid flow. In such a heterogeneous pore network, the flow pathways and the interaction between the fluid and solid matrix are complex. The interaction between the flowing fluid and the shale matrix, particularly under varying pressure and temperature conditions, can alter the pore structure dynamically (Yang et al., 2022). | |
| **Surface effects (diffusion and sorption)** | Shales, particularly organic-rich shales, exhibit significant adsorption of gas molecules onto pore wall surfaces. This adsorption, coupled with surface diffusion, where adsorbed molecules move along pore wall surfaces, can contribute to overall flow in a manner that does not align with the Darcy's law (J. Wang et al., 2016; Wu et al., 2016). | Tight sandstones exhibit significant microscopic heterogeneity and anisotropy in their pore structure. This variability can lead to uneven distribution of fluid flow, thereby contributing to deviations from the Darcy's law. The complex and tortuous pathways and large surface areas in tight sandstones can lead to increased flow resistance and contribute to nonlinear flow behaviors (Lai et al., 2018; Ma et al., 2020). |
| **Viscoelastic flow in shale matrix** | The flow within shales and tight sandstones may exhibit viscoelastic characteristics due to the interaction between the fluid and the flexible organic matter within the rock matrix (Bernadiner & Protopapas, 1994; Swartzendruber, 1962a, 1962b). This interaction can create a flow behavior that differs from the linear relationship postulated by the Darcy's law. | The surface roughness and wettability of the pore walls in tight sandstones can influence the fluid flow behavior by altering the fluid-solid interactions, thereby affecting the flow dynamics at the microscale. Organic matter and clay minerals in tight sandstones can exhibit significant adsorption and desorption of gases. These processes can affect the available pore space for flow and thus introduce |
| | | additional resistance to flow (Ghanizadeh et al., 2014; Lai et al., 2018). |
| **Electrokinetic effects** | In shales and sandstones, where electrochemical interactions are significant due to the presence of charged clay particles and saline pore water, electrokinetic effects can influence fluid flow. These effects arise from the interaction of the electrical double layers adjacent to solid surfaces, thereby causing electro-viscous effect and potentially contributing to the pre-Darcy flow behavior (Ren et al., 2001; Swartzendruber, 1962a, 1962b). | |

**Table 2.** Mechanisms causing pre-Darcy flow in clays.

| | **Clay** |
|---|---|
| **Threshold pressure gradient** | Clay often exhibits a threshold pressure gradient, which is the minimum hydraulic pressure gradient necessary to initiate fluid flow. This is due to the strong capillary forces and the electrostatic interactions within the fine pores of the clay matrix. Fluid movement only occurs once the applied hydraulic pressure gradient overcomes these forces (Liu & Birkholzer, 2012; Miller & Low, 1963). |
| **Microstructural constraints** | The microstructure of clay includes small and often tortuous pore channels and large surface areas. These channels impose significant resistance to fluid flow, thereby contributing to non-linear flow characteristics, especially at low velocities (Yang et al., 2022). |
| **Electrochemical interactions** | Clays are composed of charged particles, which interact with the ionic composition of the pore fluid. These electrochemical interactions can influence fluid movement, leading to behaviors that do not conform to the Darcy's law (Swartzendruber, 1962a, 1962b). For instance, the development of electrical double layers on clay particle surfaces causes electro-viscous effects and can impact fluid flow through the clay matrix (Ghanizadeh et al., 2014). |
| **Swelling and shrinkage** | Clays have the ability to swell when in contact with water and shrink upon drying. This swelling-shrinkage behavior can dynamically alter the pore structure, thereby impacting the fluid flow pathways and leading to time-dependent changes in permeability and flow behaviors (Whittaker et al., 2020). |
| **Viscous and non-viscous flows** | Flow through clays can be viscous or non-viscous. The viscous flow adheres to the conventional no-slip boundary condition, whereas the non-viscous flow might involve slip at the pore walls, particularly at low pressures or with non-aqueous fluids (Kutiĺek, 1972; Ren et al., 2001; Swartzendruber, 1962b). |
| **Pore clogging and unclogging** | In certain scenarios, especially in environmental engineering contexts involving filtration or contamination, clay pores can get clogged with particulate matter, which can be unclogged by shear forces under higher hydraulic pressure gradients. This dynamic process affects the flow characteristics in a non-linear manner (Wang et al., 2017; Y. Zhang et al., 2020). |
| **Surface diffusion and osmotic flow** | In some cases, surface diffusion, where adsorbed water molecules move along the surfaces of clay particles, can contribute to overall mass transport. Additionally, osmotic flow driven by concentration gradients across semi-permeable clay layers can also occur, contributing to the overall flow behavior (Li, 2001; Olsen, 1972; Whittaker et al., 2023). |

In summary, pre-Darcy flow in shales, tight sandstones, and clays involves complex interplays between microscale physical and chemical processes that challenge the classical understanding of fluid flow in porous media. In shales, these mechanisms regulate effective resource extraction, such as shale oil and gas recovery. Similarly, in tight sandstone reservoirs, a comprehensive understanding is critical to hydrocarbon extraction and groundwater management. In clays, these mechanisms are crucial in managing water flow in contexts relevant to civil, environmental, and geotechnical engineering, such as in foundations, embankments, waste contamination management, and landfills. The pre-Darcy flow behaviors in these materials require specialized modeling approaches and should be taken into account in engineering designs and environmental assessments.

## 4. Experiments on Pre-Darcy Flow

Experimental measurements of pre-Darcy flow properties are extremely challenging because it usually involves measuring ultra-low flow rates, pressure gradients, and permeability. High-precision equipment is needed in such experiments. The mechanisms that cause the nonlinear deviations from conventional linear Darcy flows in different low-permeability geomaterials play important roles in the measurements. Therefore, pre-Darcy experimental data is highly limited. The measurement of the porous medium's permeability is essential because this intrinsic property determines the onset of the pre-Darcy flow regardless of the mechanisms that cause the nonlinear flow behavior. In clayey porous media, as described before, the swelling of clays and the complicated solid-liquid interactions hinder the smooth flow of fluids. In shales, tight sandstones, and compacted sand packs, the structural complexity, such as pore structures, pore size distribution, pore connectivity, tortuosity, and the presence of clays and micro-fractures, create a complex network for fluid transport. This complexity necessitates a higher energy threshold, in the form of a pressure gradient, to overcome the inherent resistance to flow. We summarized the major experimental equipment and the underlying mechanisms in the following sections.

### 4.1. Overview of conventional and advanced experimental methods

In general, measurements of fluid flow in porous media, especially in geological porous media, require equipment consisting of a sample holder, a reservoir providing the injection fluid, gauges measuring pressures and flow rates, and energy that drives the injection fluid to flow through the sample in the sample holder. With the effluent volume measured with a stopwatch, one can calculate the flow rate through the sample, based on which and the pressure gradient the Darcy's law is used to calculate the permeability of the sample. In the modern petroleum industry, core-flooding experiments are designed based on this principle and have been widely applied for years to measure rock properties, fluid flow properties, hydrocarbon recovery potentials, etc. Core-flooding experiment is also the primary method in measuring pre-Darcy flow properties. **Figure 2** illustrates the schematic plot of a typical core-flooding equipment setup that can measure sample permeability, hydraulic pressure gradient across the sample, and flow rate under elevated temperatures and pressure conditions; the geomaterial sample in the core holder can be either a rock sample or consolidated soil sample. Note that this core-flooding equipment is designed for steady-state measurements with respect to single-phase fluid injections. To achieve two-phase fluid injections, unsteady-state depressurization is commonly recommended, and relative permeability curves for two-phase flows in porous media are usually obtained through this method. Cui et al. (Cui et al., 2008) designed an unsaturated infiltration experimental setup using the instantaneous profile method with samples that were pre-equilibrated at fixed relative humidity; several relative humidity sensors were placed across the holder to monitor the relative humidity changes across the whole sample. This setup worked well in measuring the unsaturated flow properties even though it was time consuming.

If the samples are shales and tight sandstones, the permeability measurement method is flexible. Numerous studies have demonstrated the permeability measurements in these geomaterials (Ghanizadeh et al., 2015; Lai et al., 2018; Li et al., 2020; Mukherjee & Vishal, 2023; Randolph et al., 1984; Tan et al., 2020; Tinni et al., 2012). However, the preparation of clay samples requires either pre-saturation or in-situ saturation, which results in different experimental procedures compared to shales and sandstones. In the literature, a common method is that dry clay samples are fully or partially saturated in equilibrium before they are moved to the sample holder for core flooding experiments. In many experiments related to measurements of the swelling pressure of clay, filling dry clay samples into the core holder is essential, and the in-situ water saturation can be obtained by the displacement of $CO_2$ followed by water injections (Busch et al., 2016; Dong et al., 2023; Thomas et al., 1968).

Direct measurements of the threshold pressure gradient is extremely difficult. The definition of the threshold pressure gradient, which is the hydraulic pressure gradient at the starting point of the fluid flow, directly leads to the challenges in experimental measurements. In other words, measurements of the instantaneous transition from a zero flow velocity to an infinitesimal positive flow velocity is nearly impossible in practice. Two prevalent methods for measuring threshold pressure gradients include the air-bubble method and the equivalent pressure gradient method. Miller and Low (Miller & Low, 1963) designed a method that detects the momentary onset of flow by monitoring the movement of an air bubble in a glass tube, of which the inner diameter is 0.6 mm. The linear movement of the air bubble could be detected at a resolution of 0.1 mm; thus, the minimum volume displacement that could be detected was around 0.00003 $cm^3$. The second common method to measure the threshold pressure gradient is to treat the hydraulic pressure gradient when the flow rate is adequately low as the threshold pressure gradient It should be noted that if the systematic error of the equipment is not negligible compared to the tiny hydraulic pressure that will be measured, the measurement is not acceptable even repeated measurements are conducted. There are still efforts that needs to be made to find the appropriate method other than the two methods discussed above.

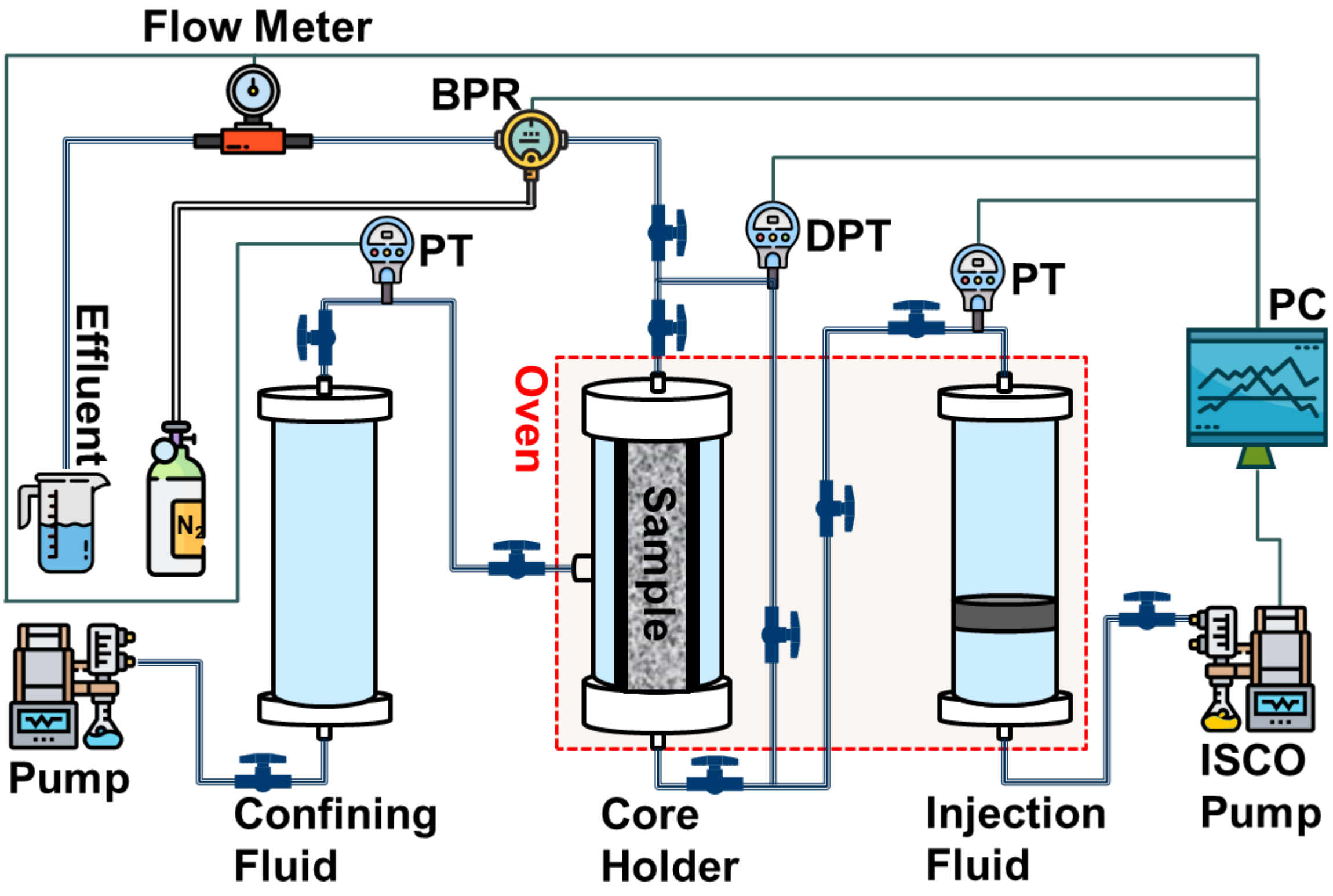

**Figure 2.** Typical core flooding equipment that can measure sample permeability, hydraulic pressure gradient across the sample, and flow rates, at elevated temperature and pressure conditions (Teng et al., 2023).

### 4.2. Experimental data

In this section, we embark on a detailed exploration of existing experimental data of pre-Darcy flow in low-permeability porous materials in the literature. Our focus is to present a comprehensive review of the measured datasets, especially the relationship between threshold pressure gradients and permeability. This analysis is crucial for understanding the variances in experimental outcomes and the factors that influence them. **Table 3** and **Table 4** illustrate all the existing laboratory-measured data with respect to pre-Darcy flow experiments, with emphasis on the geomaterial, testing fluid, experimental temperatures, permeability, and threshold pressure gradient. Particularly, Table 3 shows the experiments that considered temperature effects, whereas Table 4 shows the experiments conducted at constant temperatures. In these tables, $T$ is experimental temperature, $k$ is permeability, and $I$ is threshold pressure gradient.

All experimental data exhibited that the threshold pressure gradient increases when the permeability of the porous medium decreases. In Table 3, several studies investigated the temperature effect on pre-Darcy flow; the results showed that the threshold pressure gradients decreases with increasing temperatures under similar permeability (Teng et al., 2023; Zeng et al., 2010). Generally, permeability is an intrinsic property of the rock and does not change with temperature if the pore structure and mineral composition are thermally stable. Wang et al. (Wang et al., 2017) conducted water flooding experiments in illite, which does not swell during hydration, Therefore, the permeability of the illite sample remained at $2.91\times10^{-17}$ $m^2$, and it was found that the threshold pressure gradient decreases with increasing temperature. In the other three studies under varying temperatures, clay, mostly montmorillonite, plays an important role because of its swelling ability under hydrated conditions. Higher temperatures reduce the viscosity of the fluid and also impact the electrochemical interactions between the clay particles and water molecules (Yang et al., 2021), resulting in complicated coupled effects on the fluid flow conditions. In many theoretical studies which will be discussed in the next section, the concept of immobile water layer is important. The decline of the threshold pressure gradient with increasing permeability is faster under a higher temperature, which can be explained by the decreased static thickness of the immobile layer under a higher temperature (Teng et al., 2023). In clays, an increase in temperature can alter the zeta potential on the surfaces of particles, thereby weakening the interaction between the particles and water (Wang et al., 2017). These results further confirmed that the concept of the immobile water layer, which combines the complicated solid-liquid interactions, works well in theoretical modeling of the nonlinear pre-Darcy flow behavior. In Tables 3 and 4, several studies (Zeng et al., 2011;

Zeng et al., 2010) measured the threshold pressure gradient and permeability in tight sandstones using hydrocarbon liquids such as simulated oil made from degassed crude oil and kerosene to adjust the viscosity of the liquid and to simulate the petroleum extraction conditions. Nonlinear pore-Darcy flows were observed in these studies.

We plotted all the existing laboratory-measured data of threshold pressure gradients and permeability in the log-log scale, as shown in **Figure 3**. Overall, the data in general follows a power-law correlation. It is noted that most data points fall within the permeability range between $10^{-19}$ $m^2$ and $10^{-14}$ $m^2$ and threshold pressure gradient range between 0.5 and 400 m/m. In the next section, we will examine the current empirical models and theoretical models which attempt to depict nonlinear pre-Darcy flow behaviors from different perspectives.

**Table 3** Laboratory-measured data of threshold pressure gradient and permeability for pre-Darcy flow experiments that considered the temperature effect.

| **References** | **Sample** | **Fluid** | $J_t$ ***or*** $J_c$ | ***T*** **(°C)** | ***k*** **(m²)** | ***I*** **(m/m)** | ***T*** **(°C)** | ***k*** **(m²)** | ***I*** **(m/m)** | ***T*** **(°C)** | ***k*** **(m²)** | ***I*** **(m/m)** |
|---|---|---|---|---|---|---|---|---|---|---|---|---|
| (Teng et al., 2023) | 90% sand 10% clay | 0.1 M NaCl | $J_t$ | 20 | 4.38E-16 | 6.77 | 55 | 1.92E-16 | 8.18 | 80 | 1.23E-16 | 8.95 |
| | | | | | 2.99E-16 | 14.20 | | 2.09E-16 | 7.11 | | 1.76E-16 | 3.67 |
| | | | | | 1.40E-16 | 38.00 | | 3.18E-16 | 2.65 | | 1.15E-16 | 11.09 |
| | | | | | 4.60E-16 | 6.74 | | 2.17E-16 | 6.67 | | 1.35E-16 | 8.28 |
| | | | | | 3.75E-16 | 6.39 | | 1.52E-16 | 12.66 | | 1.61E-16 | 6.57 |
| | | | | | 3.39E-16 | 11.69 | | 2.15E-16 | 4.61 | | 2.34E-16 | 1.47 |
| | | | | | 2.65E-16 | 16.96 | | 1.67E-16 | 9.20 | | 1.64E-16 | 5.66 |
| (Wang et al., 2017) | Illite | Water | $J_c$ | 25 | 2.91E-17 | 62.39 | 50 | 2.91E-17 | 9.15 | | | |
| | | | | 35 | | 19.39 | 60 | | 2.49 | | | |
| (Miller & Low, 1963) | Na Clay | Water | $J_t$ | 10 | 8.27E-19 | 203.16 | 15 | 5.83E-19 | 152.44 | | | |
| | Li Clay | | | | 1.20E-17 | 182.28 | 20 | 4.11E-18 | 122.24 | | | |
| (Wang et al., 2022) | 30% sand 70% clay | Water | $J_c$ | 60 | 7.52E-18 | 2.27 | 40 | 7.14E-18 | 36.80 | | | |
| | | | | 50 | 7.34E-18 | 14.88 | 30 | 4.22E-18 | 57.39 | | | |
| (Zeng et al., 2010) | Sandstone | Crude oil mixed with kerosene. | $J_c$ | 90 | 2.80E-16 | 2021.41 | 90 | 4.53E-15 | 220.18 | 70 | 4.69E-15 | 464.83 |
| | | | | | 6.20E-16 | 778.80 | | 5.33E-15 | 140.67 | | 7.62E-15 | 258.92 |
| | | | | | 8.50E-16 | 548.42 | 70 | 8.20E-16 | 1391.44 | | | |
| | | | | | 1.01E-15 | 800.20 | | 1.42E-15 | 1076.45 | | | |

**Table 4** Laboratory-measured data of threshold pressure gradient and permeability for pre-Darcy flow experiments without considering the temperature effect.

| References | Sample | Fluid | $J_t$ or $J_c$ | $T$ (°C) | $k$ (m²) | $I$ (m/m) | References | Sample | Fluid | $J_t$ or $J_c$ | $T$ (°C) | $k$ (m²) | $I$ (m/m) |
|---|---|---|---|---|---|---|---|---|---|---|---|---|---|
| (Yang et al., 2022) | Sandstone | Water | $J_t$ | 30 | 6.56E-18 | 129.72 | (S. Wang et al., 2016) | Clay | Water | $J_c$ | 25 | 1.04E-17 | 2.66 |
| | | | | | 3.42E-18 | 240.21 | | | | | | 8.05E-18 | 27.20 |
| | | | | | 4.84E-15 | 1.35 | | | | | | 7.17E-18 | 30.98 |
| | | | | | 6.57E-17 | 26.90 | | | | | | 5.05E-18 | 40.19 |
| (Yang & Yu, 2020) | Shale C01 | Water | $J_t$ | 30 | 2.52E-18 | 1306.76 | | | | | | 5.51E-18 | 46.51 |
| | Shale C02 | | | | 3.48E-19 | 3439.85 | | | | | | 4.24E-18 | 57.46 |
| | Shale C03 | | | | 2.82E-18 | 1066.55 | | | | | | 3.56E-18 | 62.34 |
| (Wang et al., 2018) | 50% sand + 50% clay | Water | $J_c$ | 25 | 4.34E-18 | 11.70 | | | | | | 2.58E-18 | 63.38 |
| | 90% sand + 10% clay | | | | 5.51E-18 | 38.90 | | | | | | 2.36E-18 | 60.38 |
| | 80% sand + 20% clay | | | | 3.77E-18 | 51.30 | | | | | | 1.98E-18 | 63.40 |
| | 70% sand + 30% clay | | | | 3.17E-18 | 61.34 | | | | | | | |
| (Zeng et al., 2011) | N/A | Crude oil | $J_t$ | 20 | 4.17E-17 | 76.46 | (Zeng et al., 2011) | N/A | Water | $J_t$ | 20 | 1.22E-16 | 194.58 |
| | | | | | 7.14E-17 | 50.19 | | | | | | 2.80E-16 | 104.83 |
| | | mixed with kerosene. | | | 1.09E-16 | 31.44 | | | | | | 3.84E-16 | 73.50 |
| | | | | | 1.21E-16 | 21.02 | | | | | | 4.63E-16 | 55.27 |
| | | | | | 2.28E-16 | 11.54 | | | | | | 5.39E-16 | 58.17 |
| | | | | | 2.68E-16 | 10.04 | | | | | | 6.24E-16 | 49.16 |
| | | | | | 2.97E-16 | 6.55 | | | | | | 1.34E-15 | 22.72 |
| | | | | | 3.54E-16 | 5.36 | | | | | | 3.12E-15 | 11.36 |
| | | | | | 7.58E-16 | 6.05 | | | Crude oil mixed with kerosene. | | | 4.17E-17 | 835.51 |
| | | | | | 1.12E-15 | 2.90 | | | | | | 1.11E-16 | 290.63 |
| | | | | | 1.38E-15 | 1.71 | | | | | | 2.85E-16 | 121.15 |
| | | | | | 2.80E-15 | 1.41 | | | | | | 4.14E-16 | 63.95 |
| | | | | | 3.22E-15 | 1.17 | | | | | | 7.62E-16 | 46.25 |
| | | Water | | | 7.89E-17 | 373.94 | | | | | | 1.12E-15 | 25.20 |
| | | | | | 1.19E-16 | 258.61 | | | | | | 2.80E-15 | 11.54 |
| (Xue-wu et al., 2011) | Sandstone | Formation water | $J_c$ | N/A | 5.40E-16 | 12.20 | (Thomas et al., 1968) | N/A | 5%-10% NaCl | $J_t$ | - | 1.79E-16 | 169.31 |
| | | | | | 3.55E-16 | 10.07 | | | | | | 7.44E-17 | 198.49 |
| | | | | | 1.99E-16 | 20.61 | | | | | | 1.51E-18 | 2105.83 |
| (Cui et al., 2008) | 70% Bentonite 30 % Sand | Water | $J_c$ | N/A | 1.21E-20 | 11949.78 | | | | | | 9.97E-19 | 1764.96 |
| | | | | | 1.03E-20 | 15173.46 | | | | | | 2.97E-19 | 3066.59 |
| | | | | | 9.39E-21 | 18663.96 | | | | | | 2.61E-19 | 2588.09 |
| | | | | | 7.80E-21 | 19888.06 | | | 10% NaCl | | | 8.83E-20 | 15658.01 |
| | | | | | 7.11E-21 | 20859.65 | | | | | | 4.48E-20 | 6755.70 |
| | | | | | 4.90E-21 | 21871.67 | | | | | | 4.00E-20 | 8854.61 |
| (Prada & Civan, 1999) | Brown sandstone | Saturated brine | $J_c$ | N/A | 6.43E-12 | 0.0057 | (Prada & Civan, 1999) | Sand pack | Saturated brine | $J_c$ | N/A | 2.75E-11 | 0.0031 |
| | Brown sandstone | | | | 4.75E-12 | 0.0094 | | Sand pack | | | | 2.32E-11 | 0.0053 |
| | Brown sandstone | | | | 4.67E-12 | 0.0136 | | Sand pack | | | | 2.35E-11 | 0.0039 |
| | Brown sandstone | | | | 3.09E-12 | 0.0114 | | Shaly sandstone | | | | 5.59E-14 | 0.5442 |
| (Zou, 1996) | Clay | Water | $J_c$ | N/A | 2.81E-16 | 3.05 | (Dubin & Moulin, 1986) | Saint Herblain Clay | Water | $J_c$ | N/A | 1.61E-16 | 14.28 |
| | | | | | 9.63E-17 | 4.05 | | | | | | 8.41E-17 | 20.26 |
| | | | | | 7.64E-17 | 5.07 | | | | | | 1.31E-17 | 24.09 |
| (Blecker, 1970) | Springerville soil | Water | $J_c$ | N/A | 2.02E-14 | 0.56 | (Blecker, 1970) | Springerville soil | Water | $J_c$ | N/A | 1.98E-15 | 1.14 |
| | | | | | 2.32E-14 | 0.93 | | Brolliar soil | | | | 1.56E-15 | 0.91 |
| | | | | | 7.27E-15 | 1.14 | | | | | | 8.56E-16 | 1.77 |
| | | | | | 8.16E-15 | 1.34 | | | | | | 5.90E-16 | 2.40 |
| | | | | | 2.38E-15 | 1.80 | | | | | | 6.49E-16 | 3.69 |
| | | | | | 2.12E-15 | 1.38 | | | | | | 1.94E-16 | 3.63 |
| (Lutz & Kemper, 1959) | Utah bentonite | 0.005% HCl | $J_c$ | N/A | 9.60E-17 | 1.00 | (Lutz & Kemper, 1959) | Wyoming bentonite | 0.05% HCl | $J_c$ | N/A | 2.77E-17 | 91.96 |
| | Halloysite | 0.005% HCl | | | 9.89E-17 | 12.37 | | Balden clay | 0.05% NaCl | | | 8.05E-18 | 54.27 |
| | Halloysite | 0.005% | | | 1.34E-16 | 19.95 | | Balden clay | 0.005% | | | 1.69E-18 | 46.20 |
| | | NaCl | | | | | | | NaCl | | | | |
| | Halloysite | 0.5% NaCl | | | 3.56E-16 | 19.04 | | Wyoming bentonite | 0.005% HCl | | | 1.91E-18 | 122.13 |
| | Halloysite | Water | | | 9.71E-16 | 26.60 | | Balden clay | Water | | | 1.17E-18 | 170.57 |
| | Utah bentonite | Water | | | 1.01E-16 | 27.00 | | Wyoming bentonite | Water | | | 7.54E-19 | 307.42 |
| | Halloysite | Water | | | 1.04E-16 | 81.07 | | Utah bentonite | 0.005% NaCl | | | 1.83E-17 | 445.14 |
| | Utah bentonite | 0.5% NaCl | | | 4.83E-17 | 83.63 | | Utah bentonite | Water | | | 7.55E-18 | 513.26 |

# 5. Modeling Pre-Darcy Flow

The development of models to describe pre-Darcy flow in low-permeability porous media has been significantly influenced by experimental findings. The classical Darcy's law fails in accurately predicting fluid movement in such porous media, especially under low-velocity conditions. Pre-Darcy flow, characterized by a nonlinear relationship between the flow velocity and pressure gradient, is particularly prevalent in low-permeability geomaterials such as clays and tight sandstones. Overall, modeling of pre-Darcy flow can be broadly subdivided into two groups: empirical models and conceptual models. The empirical models describe relationship between the threshold pressure gradient ($I$) and permeability ($k$). Here we use $I$ to represent the threshold pressure gradient, which is determined by measuring either the true threshold pressure gradient ($J_t$) or the extrapolated threshold pressure gradient ($J_c$). The conceptual models are more complicated because the solid-fluid interactions and continuum flow are usually taken into account to build the fluid flow model in either a single nanotube or a pack of tubes. Both types of models can describe the nonlinear relation well. In this paper, we provide a brief overview of the models that have successfully fitted experimental data.

## 5.1. Empirical models

A well-established empirical model is the power-law model (Liu & Birkholzer, 2012), which was used to fit experimental data from seven studies (Blecker, 1970; Cui et al., 2008; Dubin & Moulin, 1986; Lutz & Kemper, 1959; Miller & Low, 1963; Xue-wu et al., 2011; Zou, 1996):

$$I = Ak^B \qquad (2)$$

where $I$ is the threshold pressure gradients (m/m) and $k$ is the permeability of the porous medium ($m^2$). The two parameters are found as $A = 4.0\times10^{-12}$ and $B = -0.78$ by fitting experimental data from the above mentioned seven studies. Note that these experimental data were all measured from clays or samples containing clays. This power-law form of the model has been proposed and developed for over half a century (Hansbo, 1960; Swartzendruber, 1962a, 1962b), which has been widely adjusted to fit measurement data in most experimental studies related to pre-Darcy flow (Prada & Civan, 1999; Siddiqui et al., 2016; Thomas et al., 1968; Wang et al., 2018; Wang et al., 2022; S. Wang et al., 2016; Wang et al., 2017; Yang et al., 2022; Zeng et al., 2011; Zeng et al., 2010). The power-law model can also describe the relationship between the flow velocity and hydraulic pressure gradient to model the nonlinear pre-Darcy flow behavior, especially under low pressure and velocity conditions:

$$U = Ai^B \qquad (3)$$

where $i$ is the hydraulic pressure gradient (Pa/m) across the sample and $U$ is the Darcy flow velocity (m/s). This is a widely used form of the modified Darcy's law to describe the nonlinear relation between the flow velocity and hydraulic pressure gradient in pre-Darcy flows (Dejam et al., 2017; Hansbo, 1960; Ing & Xiaoyan, 2002). Dejam et al. (Dejam et al., 2017) applied this equation to the continuity equation to obtain nonlinear diffusivity equations, of which the analytical solutions were derived using a generalized Boltzmann transformation technique. They modeled the one-dimensional unsteady flow of a slightly compressible fluid in homogeneous porous media under linear and radial pre-Darcy flow conditions, which provided a basis to the pressure transient behavior in low-permeability petroleum reservoirs. Other studies applied this form of the model with an exponential parameter that accounted for nonlinear behaviors of the flow velocity, and good agreements with experimental data were achieved. Some adjustments were made based on Equation 3, such as separating the hydraulic gradient into two parts for before and after the threshold pressure gradient (Yang et al., 2022) or modifying the classical Darcy's law with an exponential parameter to obtain nonlinearity (Wang & Sheng, 2017).

(Wu et al., 2023) also applied this type of power-law model (Hansbo, 1997) to their three-dimensional analytical solutions of a grouted subsea tunnel face, resulting in better modeling performance. The exponential parameter was selected from 1 to 3 based on previous studies (Dejam et al., 2017; Dubin & Moulin, 1986; Hansbo, 2001; Soni et al., 1978; Y. Zhang et al., 2020), and the value of *i* was tested from 0 through 30. This study elucidates the importance of accounting for nonlinear pre-Darcy flow behaviors in geotechnical engineering.

Wang and Sheng (Wang & Sheng, 2017) argued the existence of the threshold pressure gradient in shale reservoirs and proposed a model without the term of threshold pressure gradient:

$$v = -\frac{k}{\mu}\nabla p\left(\frac{1}{1+a\mathrm{e}^{-b|\nabla p|}}\right) \tag{4}$$

where the $v$ is the flow velocity (m/s), $k$ is the permeability (m$^2$), $\mu$ is the fluid viscosity (Pa·s), $\nabla p$ is the pressure gradient (Pa/m), and $a$ and $b$ are two positive dimensionless coefficients. When a=0, equation 4 becomes the classical Darcy's law. To obtain $a$ and $b$, polynomial fittings of the experimental data of (Xue-wu et al., 2011) to the model were conducted. Yan et al. (Yan et al., 2024) applied Equation 4 to their analytical solution for the productivity prediction of horizontal wells with multi-stage fractures in a shale formation and achieved a lower relative error compared to the widely used semi-analytical solution.

To further examine the power-law model, we fitted the entire measurement data of the threshold pressure gradient and permeability to Equation 2, as shown in **Figure 3**. The fitted parameters of $A$ and $B$ are $5.74\times10^{-9}$ and −0.61, respectively, with $R^2$=0.92. Liu and Birkholzer (Liu & Birkholzer, 2012) found $A = 4.0\times10^{-12}$ and $B = -0.78$ by fitting the same model to a subset of the dataset. The relatively large discrepancy in the values of $A$ indicates that the power-law model is sensitive to the dataset that is used to fit the parameters. Nevertheless, the power-law model provides a simple and straightforward tool for predicting the threshold pressure gradient based on the porous medium's permeability.

In many subsurface conditions, temperature is a key factor influencing the fluid flow in geomaterials. The above mentioned empirical power-law model cannot quantify the influence of temperature on pre-Darcy flow behaviors. Liu (Liu, 2014) updated the power-law model with two extra parameters, the reference viscosity and actual viscosity:

$$I = Ak^B \left( \frac{\mu_{ref}}{\mu} \right)^B \qquad (5)$$

where $\mu_{ref}$ is the reference viscosity (Pa·s) at room temperature (e.g., 25°C), and $\mu$ is the actual fluid viscosity at the target temperature. We fitted the laboratory data of threshold pressure gradient and permeability at elevated temperatures measured by (Teng et al., 2023) and (Zeng et al., 2010) to Equations 2 and 5; the results were plotted in **Figure 4.** We selected 20°C as the reference temperature for the dataset in (Teng et al., 2023) and 70°C as the reference temperature for the dataset in (Zeng et al., 2010). Therefore, the fitting of the data at the reference temperature was based on Equation 2. We obtained $A = 2.79\times10^{-19}$ and $B = -1.52$ for (Teng et al., 2023), and $A = 9.80\times10^{-5}$ and $B = -7.41$ for (Zeng et al., 2010). The fitted values of $A$ and $B$ were then used in Equation 5 to predict the threshold pressure gradients at higher temperatures. The data fitting showed that Equation 5 did a good job in describing the overall effect of temperature on the threshold pressure gradient. However, it failed to predict the rapid decline of the threshold pressure gradient at higher temperatures, as shown in the dataset of Teng et al. This suggests that the empirical power-law model can capture pre-Darcy flow behaviors to some extent, but the application of this type of empirical models has its limitations.

In the next section, we will discuss the theoretical models that account for more fundamental mechanisms that regulate pre-Darcy flows in low-permeability porous media.

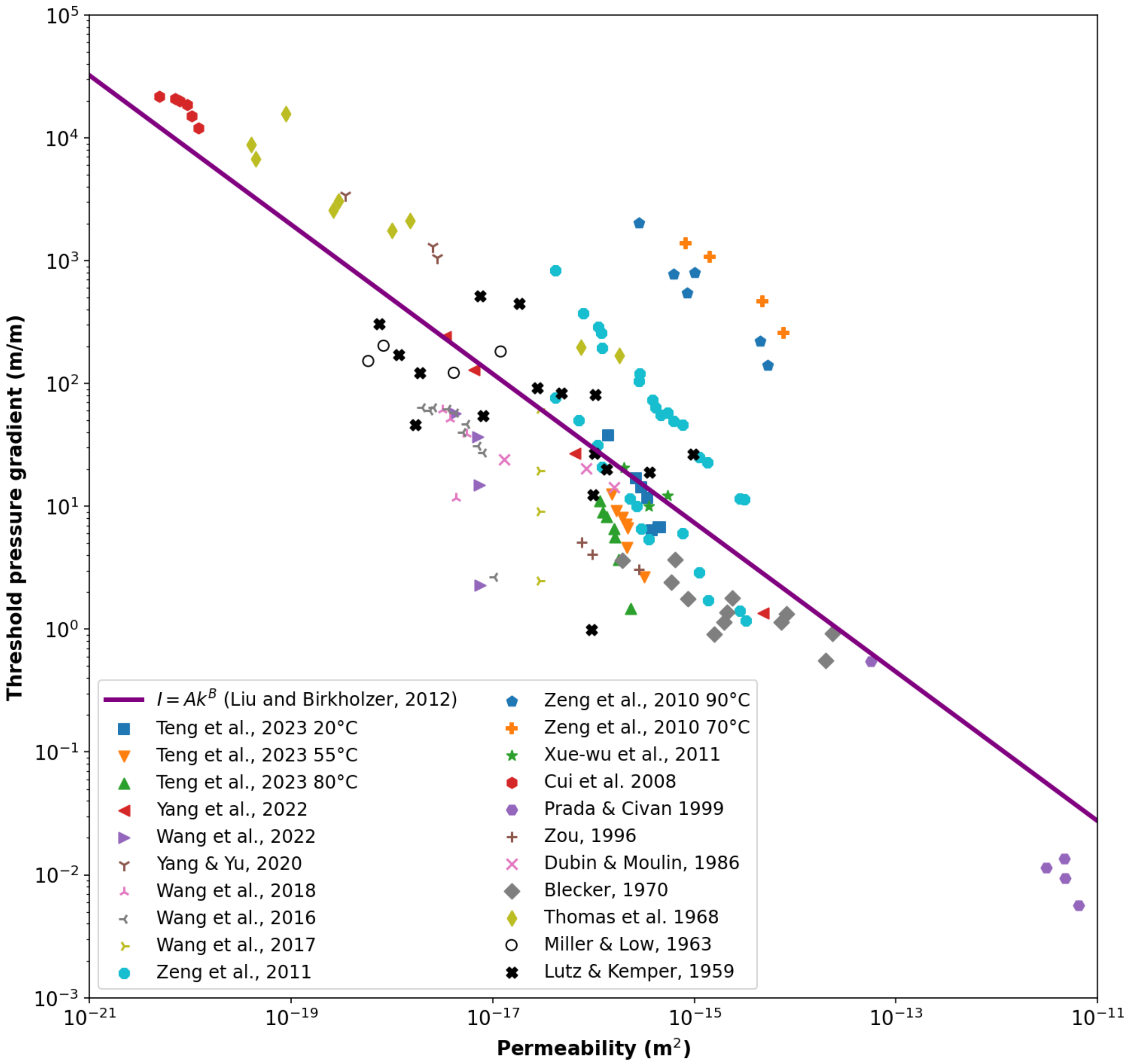

**Figure 3** Existing laboratory measurement data of the threshold pressure gradient and permeability, as well as power-law data fitting.

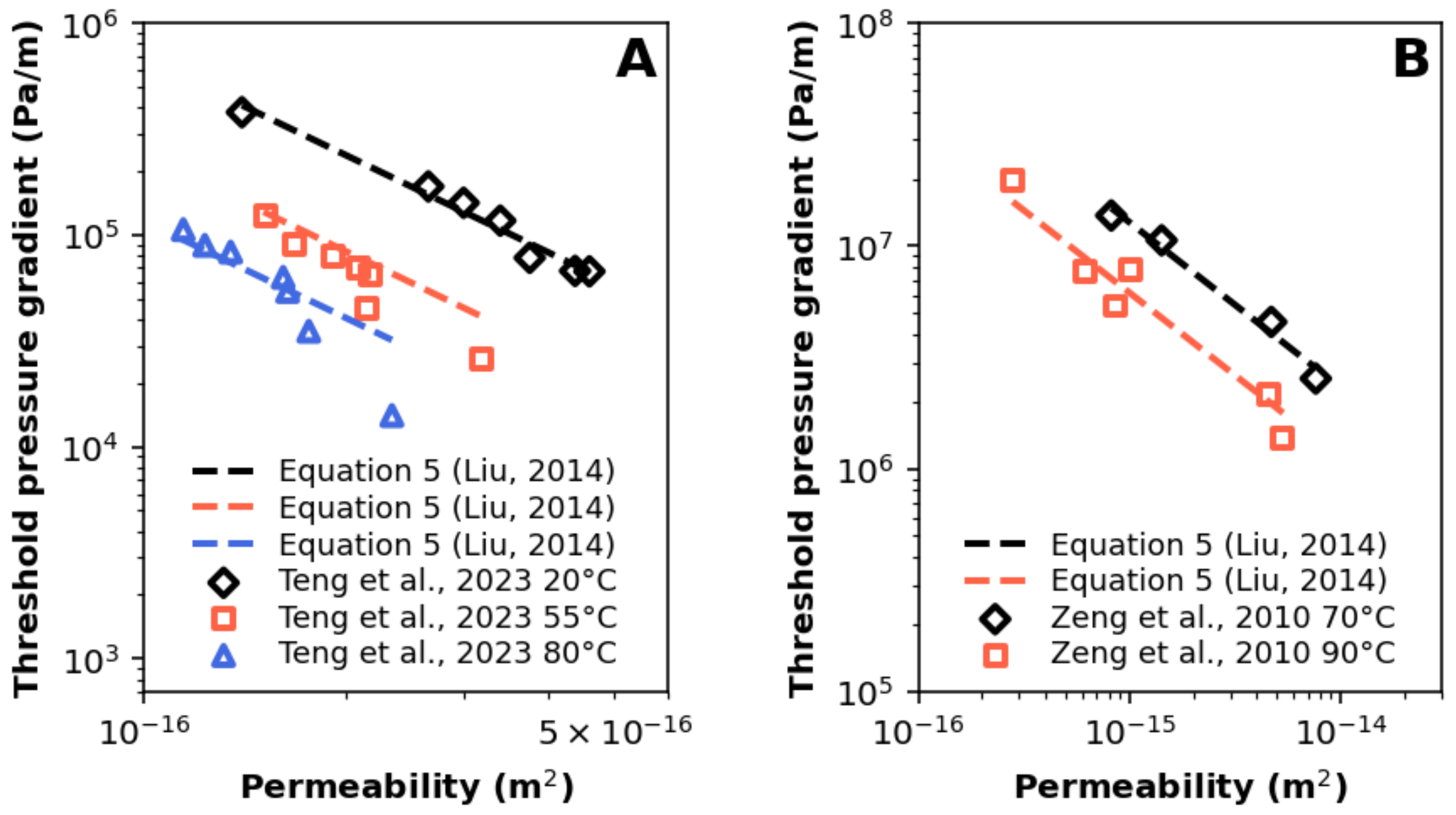

**Figure 4.** Temperature-dependent data of the threshold pressure gradient and permeability fitted to Equation 5. **A:** Data from (Teng et al., 2023). **B:** Data from (Zeng et al., 2010).

### 5.2.Theoretical models

Miller and Low (Miller & Low, 1963) measured the threshold pressure gradient against the permeability of swelling bentonite at two temperatures; they proposed that the complex solid-liquid interactions play an important role in pre-Darcy flows. Below the critical hydraulic pressure gradient, which is actually the extrapolated threshold pressure gradient in their study, water is immobilized so the pressure gradient needs to overcome a certain amount of binding energy to trigger the flow. Miller and Low hypothesized that this binding energy depends on the distance to the solid surface. Based on this hypothesis, some theoretical models were developed based on the relation between the pressure gradient and thickness of an immobile water layer on solid surface.

Chen (Chen, 2019) assumed that the thickness of the immobile water layer decreases exponentially with the increase of the flow velocity gradient on the pore wall. Based on this hypothesis, Chen developed a two-parameter model to describe the relationship between the true threshold pressure gradient and the thickness of the immobile water layer:

$$J_t = \begin{cases} \dfrac{-2\mu}{aR}\ln(\dfrac{R}{h_0}) & \text{if } h_0 > R \\ 0 & \text{if } h_0 \le R \end{cases} \quad (6)$$

where $h_0$ is the static thickness of the immobile water layer (m), $R$ is the circular tube radius (m) which is related to the apparent permeability of the porous medium, $\mu$ is the fluid viscosity (Pa·s), and $a$ is the characteristic time (s). The values of the two parameters, $h_0$ and $a$, under a particular temperature can be determined by fitting the model (i.e., Equation 6) to experimental measurement data using the nonlinear least square fitting method. In this two-parameter model, the thickness of the immobile water layer as a function of the pressure gradient can be described as:

$$h = h_0 \exp\left(-\frac{aR}{2\mu}\left|\frac{\partial p}{\partial x}\right|\right) \quad (7)$$

where $h$ is the thickness of the immobile water layer (m). In this way, one can substitute Equation 7 into the empirical equation describing the apparent permeability of a pack of circular tubes with a porosity $\phi$, written as $k_p = \phi(R-h)^2/8$. The apparent permeability can then be substituted into the classic Darcy's law to describe the relationship between the Darcy flow velocity and hydraulic pressure gradient (Teng et al., 2023):

$$U = -\frac{\phi}{8\mu}\left[R - h_0 \exp\left(-\frac{aR}{2\mu}\left|\frac{\partial p}{\partial x}\right|\right)\right]^2 \frac{\partial p}{\partial x} \quad (8)$$

We fitted the model of Equation 6 to the laboratory measurement data from Teng et al. (Teng et al., 2023) and Zeng at al. (Zeng et al., 2010), as shown in **Figure 5**. The data selection is the same as the modeling of Equation 5 and Figure 4. As the Figure 5 shows, Equation 6 predicts the temperature-dependent threshold pressure gradients well for all the data of (Teng et al., 2023) and the 70°C data of (Zeng et al., 2010). For the 90°C data in Figure 5B, there are deviations in the region of higher permeability, which could be because of the hydrocarbon fluid and the sandstone rock properties. To simulate the production of hydrocarbons, their sandstones cores were initially saturated with water, and then the measurements of flow properties were conducted when the injected oil displaced water. The solid-liquid interactions between oil and sandstone are not as heavier as solid-liquid interactions between water and swelling clays and the two-phase flow is more complicated. Equation 6 also takes into account the faster decline of threshold pressure gradient with increasing permeability at higher temperature.

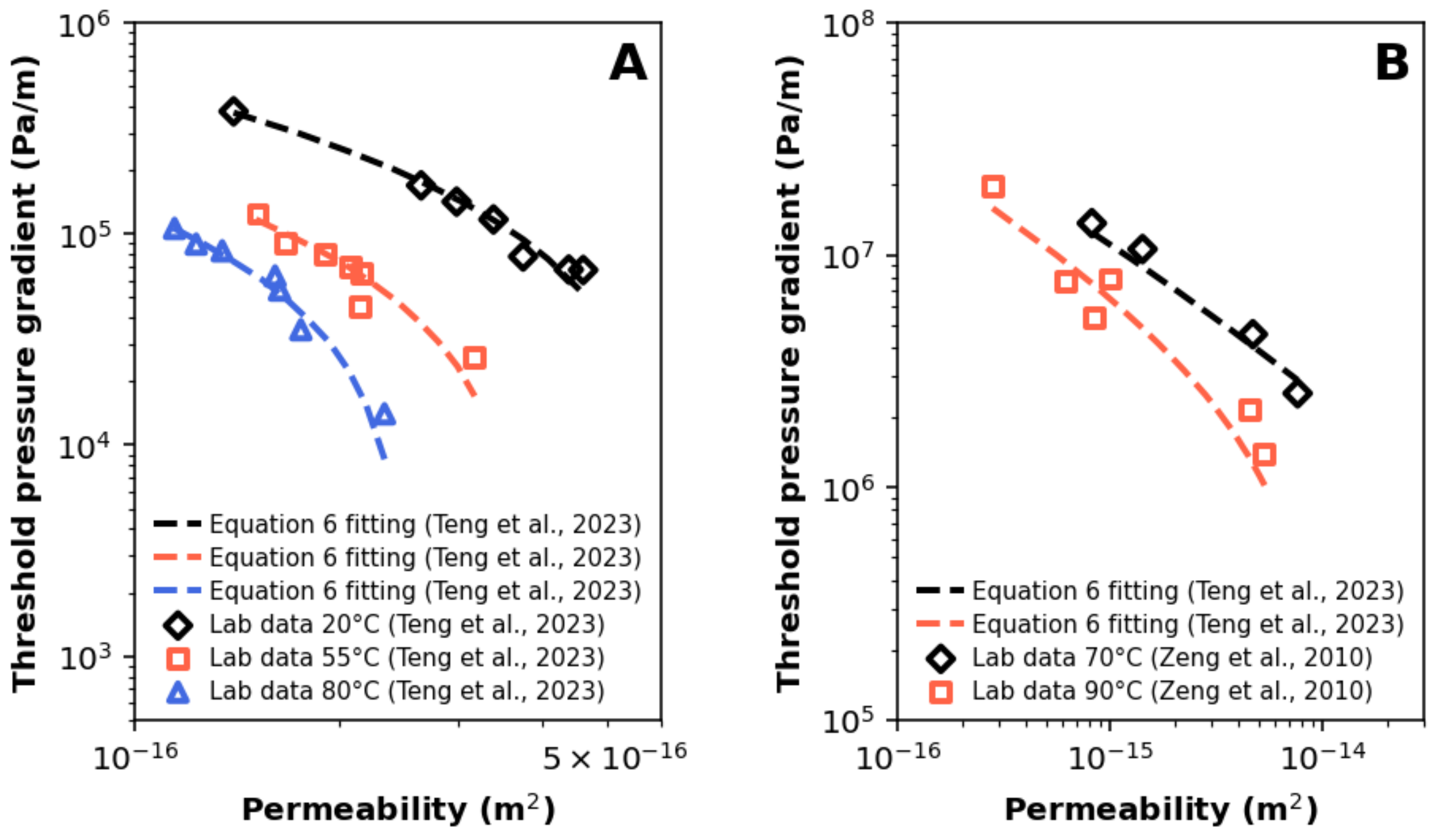

**Figure 5.** Fitting Equation 6 to laboratory measurements from A) (Teng et al., 2023), and B) (Zeng et al., 2010).

Jiang et al. (Jiang et al., 2012) developed a nonlinear model for seepage flows in low-permeability oil reservoirs based on a capillary flow model and the boundary layer effect:

$$v = -\frac{k}{\mu}\left(1 - \frac{c_1}{|\nabla p| - c_2}\right)\nabla p \qquad (9)$$

where $v$ is flow velocity (m/s), $k$ is permeability ($m^2$), $\mu$ is fluid viscosity (Pa·s), $\nabla p$ is pressure gradient (Pa/m), and $c_1$ and $c_2$ are two parameters to characterize the nonlinearity of the system. If $c_1$=0, Equation 9 reduces to the classic Darcy's law. If $c_2$=0, Equation 9 reduces to the quasi-linear flow model. The threshold pressure gradient is the sum of $c_1$ and $c_2$. Different from the previously mentioned empirical models, this model is based on two parameters combined with the reciprocal of $\nabla p$ to characterize threshold pressure gradient and the nonlinearity. Wang et al. (Wang et al., 2020) modified this model by consolidating the two parameters into one parameter to investigate pre-Darcy flow behaviors in $CO_2$ huff-n-puff operations in tight oil reservoirs.

Similarly, Xiong et al. (Xiong et al., 2017) also developed a model based on the capillary flow model and boundary layer effect. They applied the exponential formulation to account for the nonlinearity:

$$v = \frac{k\left(1 - \delta_{\mathrm{D}} \mathrm{e}^{-c_{\phi}|\nabla p|}\right)^4}{\mu}\nabla p \qquad (10)$$

where $v$ is flow velocity (m/s), $k$ is permeability ($m^2$), $\mu$ is fluid viscosity (Pa·s), $\nabla p$ is pressure gradient (Pa/m), $\delta_D$ is the dimensionless number, equal to the ratio of the boundary layer thickness to the tube radius, and $c_\varphi$ is a regression parameter which is related to both the fluid and properties of the porous medium, such as surface wettability and pore structures.

We selected a dataset from (Teng et al., 2023) to examine the theoretical models described above. This dataset was measured at 20°C, covering both the pre-Darcy and Darcy flow regimes. The threshold pressure gradient was measured as $1.42\times10^{5}$ Pa/m, and the permeability of the clay sample was measured as $2.99\times10^{-16}$ $m^2$. The examination is shown in **Figure 6**. The fitting of Equation 3 is easy because of its simple formulation, but the predicted flow velocity at the threshold pressure gradient is greater than zero. We calculated the parameters of $a$ and $b$ in Equation 4 based on the polynomial equations in (Wang & Sheng, 2017). Because Equation 4 abandoned the concept of threshold pressure gradient, the flow velocity started from a zero hydraulic pressure gradient. Equation 8 takes full account of the threshold pressure gradient and predicts the flow velocity well with increasing hydraulic pressure gradient. Equation 9 has two constants, $c_1$ and $c_2$, and the sum of them is the threshold pressure gradient. We applied nonlinear least square fitting to determine the values of $c_1$ and $c_2$. The fitted curve of Equation 9 shows that the flow velocity at the threshold pressure gradient is also greater than zero. Equation 10 has two parameters, $\delta_D$ and $c_\varphi$, and their values were determined using the thickness of the immobile water layer and the equivalent radius of a bundle of tubes (Teng et al., 2023). The calculated flow velocity at the threshold pressure gradient based on this model is lower than zero.

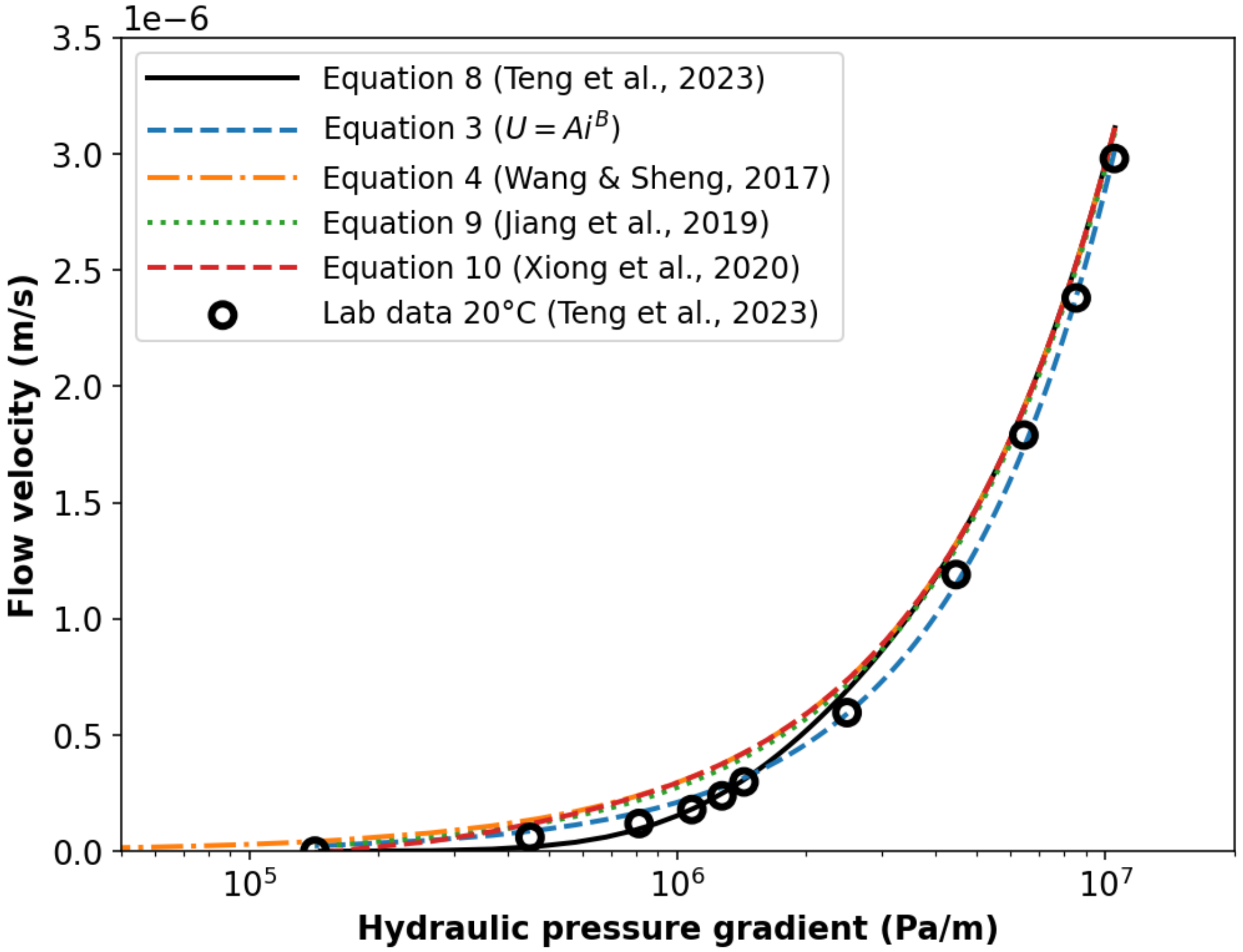

**Figure 6.** Examination of various theoretical models using the laboratory measurement data at 20°C from Teng et al. (Teng et al., 2023). This dataset covers both the pre-Darcy and Darcy flow regimes, with the threshold pressure gradient being $1.42\times10^5$ Pa/m and the permeability of the clay sample being $2.99\times10^{-16}$ $m^2$.

These theoretical models aim to provide a more accurate and predictive understanding of pre-Darcy flow behaviors in low-permeability porous media. However, both empirical and theoretical models have their advantages and limitations. Empirical models are simple and can fit a wide range of experimental data. However, the values of the fitted parameters may fluctuate significantly due to the changes in the dataset, and the models are unable to account for the temperature effect. Theoretical models, to some extent, account for fundamental mechanisms that influence and regulate pre-Darcy flow behaviors. However, they are more complicated and sometimes contain more undetermined parameters, of which the values may require nonlinear lease square fitting to determine.

## 6. Challenges and Future Directions

### 6.1.Challenges in experimental investigations

Conducting experiments on pre-Darcy flow, especially in low-permeability porous media, is subject to several challenges:

- **Measurement sensitivity and accuracy**: Pre-Darcy flow often involves extremely low flow velocities and small pressure gradients, thereby requiring highly sensitive and accurate measurement equipment. This sensitivity is essential to detect subtle changes in flow properties and to accurately characterize the non-linear relationship between the flow velocity and pressure gradient.
- **Sample heterogeneity and representation**: Low-permeability geomaterials, including clays, tight sandstones, and shales, often exhibit heterogeneity in their pore structure and mineral composition. Ensuring that the samples used in the laboratory are representative of the large-scale geological formations can be challenging. This heterogeneity can lead to variability in experimental results, thereby resulting in challenges in the interpretation of data.
- **Control of experimental conditions**: Precisely controlling and maintaining experimental conditions, such as temperature, pressure, and fluid properties, are crucial for reliable experiments. Fluctuations in these conditions can significantly impact flow behaviors in low-permeability porous media, especially when the hydraulic pressure gradient is low.

- **Determination of the threshold pressure gradient**: Identifying the threshold pressure gradient (i.e., the point at which the flow occurs and enters the pre-Darcy regime) is difficult. It requires careful experimentation and analysis, because this transition is not always easy to capture and can be influenced by multiple factors, such as the pore structure and fluid characteristics.
- **Fluid-solid interactions**: Understanding and accounting for the complex interactions between the fluid and the porous matrix, such as adsorption, chemical reactions, and mechanical effects, is challenging but essential for accurately modeling of pre-Darcy flows.
- **Data interpretation and model validation**: Interpreting experimental data to develop or validate flow models is complicated, especially given the nonlinear properties related to pre-Darcy flows. Models need to account for a range of factors and must be validated against diverse sets of experimental data.
- **Reproducibility and consistency**: Achieving reproducibility in experiments and ensuring consistency across different studies are crucial for the development of universally accepted models. Variability in experimental setups, procedures, and sample characteristics can lead to inconsistent results, thereby hindering the advancement of the fundamental understanding in pre-Darcy flows.
- **Costs and resources**: Conducting well-controlled laboratory experiments with respect to pre-Darcy flows often requires sophisticated equipment, specialized facilities, and significant time investment, thereby making such studies resource-intensive. For example, one unsaturated flow experiment conducted by (Cui et al., 2008) took thousands of hours.
- **Complexity of porous media**: The complexity of porous media, such as pore size distribution, tortuosity, and anisotropy, adds additional layers of complexity to experimental investigations, thereby requiring advanced techniques and methodologies to unravel these complexities.

Addressing these challenges is critical to advancing our understanding of pre-Darcy flow behaviors and developing reliable models that can be applied to various geotechnical, hydrological, and petroleum engineering contexts.

### 6.2.Future directions

Future directions of the research on pre-Darcy flows in low-permeability porous media are likely to be focused on several key areas, based on current understanding and emerging challenges in the field:

- **Advanced modeling and simulation techniques**: Enhanced numerical models that accurately simulate pre-Darcy flow behaviors are needed. These models would incorporate complicated factors such as pore structure variability, stress-dependency of permeability, and non-linear flow characteristics. The development of high-fidelity simulation tools capable of handling these complexities is crucial.
- **Experimental investigations**: There is an urgent need for more sophisticated experiments that can validate and refine theoretical models. This includes laboratory experiments under well-controlled conditions to better understand the microscale mechanisms regulating pre-Darcy flows and field-scale studies to capture real-world complexities.
- **Integration of multi-scale data**: Integration of data from various scales, ranging from nanoscale pore structure analyses to field-scale observations, could provide a more comprehensive understanding of pre-Darcy flow behaviors. This multi-scale approach would help in developing models that are both theoretically sound and practically applicable.
- **Materials characterization**: Understanding the impact of different geomaterials on pre-

Darcy flow behaviors is essential. This involves studies of various types of low-permeability geomaterials including shales, tight sandstones, and clays to determine how the specific material properties influence flow behaviors, such as the swelling hysteresis of smectites that influences the flow behaviors in clay during hydration and dehydration processes.

- **Impact of external factors**: Investigations of the effects of external factors such as temperature, chemical interactions, and mechanical stress on pre-Darcy flows could lead to better predictions of flow behaviors under varying environmental and operational conditions.
- **Coupled processes**: Delving deeper into the coupled thermal-hydrological-mechanical-chemical (THMC) processes that affect flow in low-permeability media is critical to optimization of various engineered processes such as geothermal energy recovery, geological carbon sequestration, and subsurface nuclear waste disposal.
- **Machine learning and advanced imaging techniques**: Utilization of machine learning and advanced imaging techniques to visualize flow conditions in low-permeability porous media and to analyze complex datasets to predict pre-Darcy flow behaviors will be particularly useful in optimizing resource extraction and managing environmental impacts.

In conclusion, future research directions in the study of pre-Darcy flows in low-permeability porous media are multidisciplinary, thereby requiring a blend of experimental, theoretical, and computational approaches. These efforts will not only advance scientific understanding but also provide practical implications to various research fields such as hydrogeology, petroleum engineering, and environmental management.

## 7. Conclusion

This comprehensive review has systematically discussed pre-Darcy flows in low-permeability porous media from various aspects, thereby underlining its significance across many research and industrial fields, such as petroleum engineering, geothermal engineering, geological carbon storage, subsurface hydrogen storage, contamination remediation in aquifers, and geological disposal of nuclear waste. Our investigation spanned from the fundamental principles of pre-Darcy flow, including its deviation from the classical Darcy's Law, to the complex mechanisms governing flow behaviors in geomaterials such as shale, tight sandstone, and clays. We delved into different experimental methodologies and challenges, thereby highlighting the equipment sensitivity and precision required in measuring ultra-low flow rates and pressure gradients, as well as the difficulty in replicating heterogeneous and complex natural systems in laboratory settings.

Furthermore, the review shed light on the theoretical and empirical modeling approaches that attempt to characterize pre-Darcy flows. These models, while offering valuable insights into pre-Darcy flows, also underscore the inherent challenges in accurately predicting flow behaviors due to the nonlinear and multifaceted nature of low-permeability porous media. The empirical models are in general easier to use due to their simple mathematic formulation but may fail to account for complicated fundamental mechanisms that regulate pre-Darcy flows. On the other hand, the theoretical models offer a more nuanced understanding of the underlying mechanisms.

The challenges In experimental Investigations and modeling underscore the urgent need for more advanced, integrated, and multidisciplinary approaches. Future research should focus on enhancing modeling techniques, conducting sophisticated experiments, integrating multi-scale data, and exploring the impacts of external factors such as temperature, chemical interactions, and mechanical stresses. The exploration of coupled THMC processes will be pivotal in many applications ranging from resource extraction to environmental management.

In conclusion, pre-Darcy flow in low-permeability porous media represents a frontier of scientific research that has practical implications to many natural and engineered processes. Advancements in this field not only deepen our fundamental understanding of subsurface fluid dynamics but also have applications to efficient and sustainable practices in hydrogeology, petroleum engineering, and environmental engineering. As we continue to uncover the complexities of pre-Darcy flows, it is imperative that our approaches should remain adaptive and integrate emerging technologies and scientific insights to address the evolving challenges in understanding and managing fluid flows in the subsurface environment.

## Declaration of Competing Interest

The authors declare that there are no conflicts of interest associated with this study.

## Data Availability

Data are available through: 10.5281/zenodo.10400616

## Acknowledgements

The authors are thankful to the financial support provided by the U.S. Department of Energy (DOE)'s Nuclear Energy University Program (NEUP) through the Award Number of DE-NE0009160.